# Coexistence of superconductivity and charge-density wave in the quasi-one-dimensional material HfTe$_3$

Saleem J. Denholme[1,*], Akinori Yukawa[1], Kohei Tsumura[1], Masanori Nagao[2], Ryuji Tamura[3], Satoshi Watauchi[2], Isao Tanaka[2], Hideaki Takayanagi[2] and Nobuaki Miyakawa[1,*]

[1]Tokyo University of Science, Department of Applied Physics, Tokyo, 125-8585, Japan
[2]Tokyo University of Science, Department of Materials Science and Technology, Tokyo, 125-8585, Japan
[3]University of Yamanashi, Interdisciplinary Graduate School of Medicine and Engineering, Tokyo, 400-8511, Japan
[*]sdenholme@rs.tus.ac.jp
[*]miyakawa@rs.tus.ac.jp

**ABSTRACT**

We present the first experimental evidence for metallicity, superconductivity (SC) and the co-existence of charge density waves (CDW) in the quasi-one-dimensional material HfTe$_3$. The existence of such phenomena is a typical characteristic of the transition metal chalcogenides however, without the application of hydrostatic pressure/chemical doping, it is rare for a material to exhibit the co-existence of both states. Materials such as HfTe$_3$ can therefore provide us with a unique insight into the relationship between these multiple ordered states. By improving on the original synthesis conditions, we have successfully synthesised single phase HfTe$_3$ and confirmed the resultant structure by performing Rietveld refinement. Using low temperature resistivity measurements, we provide the first experimental evidence of SC at ~1.4 K as well as a resistive anomaly indicative of a CDW formation at ~82 K. By the application of hydrostatic-pressure, the resistivity anomaly shifts to higher temperature. The results show that HfTe$_3$ is a promising new material to help study the relationship between SC and CDW.

**Introduction**

The coexistence of superconductivity (SC) and charge- and/or spin-density waves (CDW and/or SDW) is fundamental to our understanding behind the mechanism of high-T$_c$ SC and is one of the most significant challenges in condensed matter physics.[1] CDWs favor low dimensionality[2] and materials such as the layered/chain-like transition metal di/trichalcogenides (MQ$_2$/Q$_3$) (where M = groups IV- VI transition metals and Q = sulfur, selenium and tellurium) have been collectively studied for such phenomena.[3] Examples include TaS$_3$[4] and NbS$_3$[5] which exhibit CDWs and TaSe$_3$ which becomes a SC below 2.1 K.[6] Studies of SC in the MQ$_2$/Q$_3$ family often support a competitive relationship between the SC and CDW states; SC can be induced/enhanced by the suppression of the CDW. This is typically achieved by the application of hydrostatic pressure such as in the case of NbSe$_2$[7] and NbSe$_3$[8] or by chemical doping for Na$_x$TaS$_2$[9] and Cu$_x$TiSe$_2$.[10] However, it is rare that the materials without chemical/physical modification exhibit the co-existence of both states. ZrTe$_3$ is a material which shows the coexistence of a CDW at ~63 K and filamentary SC at 2K[11] as does NbSe$_2$.[3] In the case of ZrTe$_3$, by the application of pressure, intercalation of Cu[12] and Ni[13] or the substitution of Se at the Te site,[14] the CDW can be suppressed and bulk SC induced at ~5K.[15] The electronic structure of ZrTe$_3$ is unique amongst the MQ$_3$ family owing to the strong contribution of the Te-Te p$_{\sigma*}$ band at the vicinity of the Fermi level,[16] therefore the inter-chain interactions affects the electronic structure as well as the physical properties. Similar cross-chain interactions are absent in other members of the MQ$_3$ family (when M = group IV transition metal and Q = S/Se).[17] Of the MTe$_3$ materials, HfTe$_3$ is the only other material expected theoretically.[18,19] There are no known reports for TiTe$_3$ nor Nb/TaTe$_3$. However, by using the reaction conditions outlined by Brattås *et al*. we found that the successful synthesis of HfTe$_3$[18,19] was irreproducible. Therefore, although theoretical band structure calculations have predicted HfTe$_3$ to be metallic[16,20,21] there is currently no experimental confirmation. As far as the authors are aware, the available experimental data for HfTe$_3$ include the original structural characterization,[18] and the determination of its basic magnetic properties (temperature-independent diamagnetism).[19] In addition, it has been recently reported by scanning tunneling spectroscopy that Hf/HfTe$_5$/HfTe$_3$ films exhibited a superconducting gap-like spectra.[22] HfTe$_3$ and ZrTe$_3$ are iso-structural materials whose features raise the possibility that HfTe$_3$ may also exhibit the coexistence of SC and CDW state. Therefore, it would be an important task to synthesize the high quality bulk compound, and to explore the aforementioned electrical phenomena.

By modifying the original synthesis conditions,[18,19] polycrystalline HfTe$_3$ samples have been successfully synthesized. The crystal structure has been analyzed using Rietveld refinement and the first experimental evidence of metallicity in this material is reported. The resistivity data exhibits an anomaly suggestive of a CDW formation at ~82 K and subsequently zero resistivity below 2K. By the application of hydrostatic pressure, the resistivity anomaly shifts to higher temperature. In addition, we note that HfTe$_3$ is highly air-sensitive, where the behaviour of ρ-*T* characteristics changes from metallic to insulating upon exposure in air (See Supplementary information).

## Results and Discussion

**Key requirements to synthesise single phase HfTe$_3$.** Suitable reaction conditions to produce single phase HfTe$_3$ crucially depend on the maximum reaction temperature.[19] During this investigation it has been found that a slow cooling rate is also a key requirement. In brief, the favoured phase was HfTe$_2$ at a higher temperature range (≥ 530 °C) and HfTe$_5$ at lower temperature regions (≤ 470 °C). As reported by Brattås *et al.*, we confirmed that the sintering condition of c.a. 500 °C certainly favours the growth of the HfTe$_3$ phase.[18] However, when rapid cooling from 500 °C (*e.g.* quenching in water) was applied[19] the majority phase became HfTe$_2$ together with unreacted tellurium. On the other hand, when slow cooling was performed (approx. -0.25 °C/h) until 470 °C after which the ampoules were cooled to room temperature at a rate of approx. -5 °C/h, then single phase HfTe$_3$ could reproducibly be synthesised. The results suggest that HfTe$_3$ primarily forms by reaction with the tellurium vapour upon cooling. If the reaction vessel is quenched, the solidification of the tellurium prevents its uptake and HfTe$_2$ becomes the preferred phase. Namely, it is found that HfTe$_3$ is the least thermodynamically stable phase within the Te-rich Hf alloys and as a result in order to inhibit the formation of trace amounts of HfTe$_2$/HfTe$_5$, it is necessary to control precisely both the sintering temperature and the cooling rate.

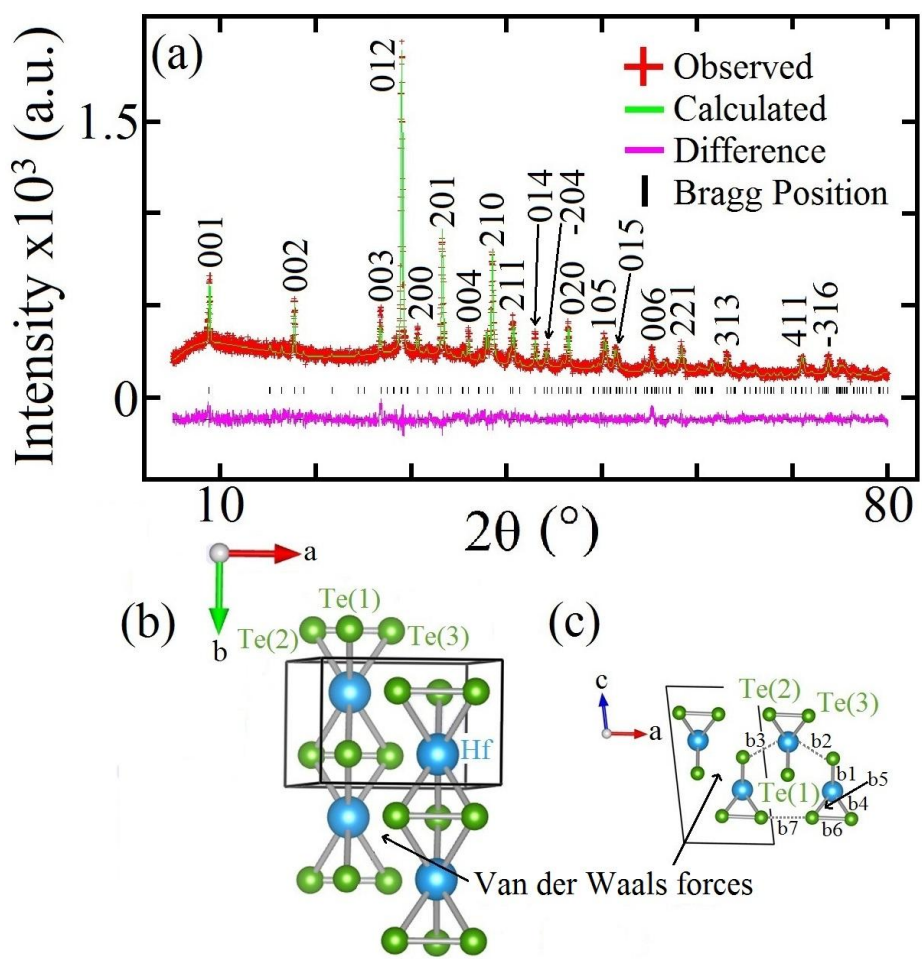

**Figure 1.** (a) Rietveld analysis of the PXRD results for HfTe$_3$. (b) Crystal structure of HfTe$_3$ revealing the anisotropic growth preferential along the *b*-axis. (c) Projection down the *b*-axis showing more clearly the separation of the chains, where the chains are weakly bonded by the Van der Waals forces. The positions of the three non-equivalent Te atoms are defined as Te(1), Te(2) and Te(3) and bond distances are indicated by b1-b7. The unit cell is indicated by the black lines.

**Crystallographic analysis.** Figure 1(a) shows the powder X-ray diffraction (PXRD) result for HfTe$_3$ together with the result of the Rietveld refinement using ZrSe$_3$ as a reference model,[23] where the result was consistent with the monoclinic crystal symmetry (space group P2$_1$/m). Figure 1(b) represents the crystal structure of HfTe$_3$ which is the pseudo-one-dimensional (1D) structure. As seen in Fig. 1(b), MQ$_6$ trigonal prismatic units propagate along the *b*-axis resulting in chain-like anisotropic crystal growth. By projection down the *b*-axis it can be clearly seen how the chains are bonded together by Van der Waals forces (see Fig. 1(c)). Reasonable values of R$_{wp}$ = 8.47%, R$_p$ = 6.60% and χ$^2$ = 1.544 were obtained. Refined lattice parameters of HfTe$_3$, *a* = 5.8797(9) Å, *b* = 3.8999(9) Å, *c* = 10.0627(3) Å agreed with the previously reported values.[19] On the other hand, the angle β = 98.38(8)° showed a slight expansion from the originally reported angle of β = 97.98°.[18] The refinement results are summarized in Table 1. It was confirmed from XRF results also confirmed that the composition ratio of our HfTe$_3$ was Hf:Te=26:74 (at%).

| Crystal system | Monoclinic | No. Observations | 7500, |
|---|---|---|---|
| Space group | P2$_1$/m (No. 11) | No. parameters | 31 |
| *a* (Å) | 5.8797(9) | R$_{wp}$ | 0.0847 |
| *b* (Å) | 3.8999(9) | R$_p$ | 0.066 |
| *c* (Å) | 10.0627(3) | Goodness of fit, χ$^2$ | 1.544 |
| β (°) | 98.38(8) | Temperature (K) | 295 K |
| V (Å$^3$) | 228.28(2) | | |
| Z | 2 | | |
| **Hf; 2*e* (x, 1/4, z)** | | **Te(1); 2*e* (x, 1/4, z)** | |
| x | 0.2590(7) | x | 0.7339(7) |
| z | 0.6881(2) | z | 0.5674(3) |
| 100 x U$_{iso}$ (Å$^2$)* | 0.484 | 100 x U$_{iso}$ (Å$^2$) | 0.484 |
| Occ | 1 | Occ | 1 |
| **Te(2); 2*e* (x, 1/4, z)** | | **Te(3); 2*e* (x, 1/4, z)** | |
| x | 0.4173(7) | x | 0.9058(6) |
| z | 0.1625(9) | z | 0.1638(3) |
| 100 x U$_{iso}$ (Å$^2$)* | 0.484 | 100 x U$_{iso}$ (Å$^2$)* | 0.484 |
| Occ | 1 | Occ | 1 |
| **Selected bond lengths (Hf-Te) (Å)** | | **Selected bond lengths (Te-Te) (Å)** | |
| Hf - Te(1)  x2  (b1)$^†$ | 3.074(6) | Te(2) - Te(3) (Å)  x1  (b6)$^†$ | 2.870(13) |
| Hf - Te(1)  x1  (b2)$^†$ | 3.106(8) | Te(2) - Te(3) (Å)  x1  (b7)$^†$ | 3.010(13) |
| Hf - Te(1)  x1  (b3)$^†$ | 3.106(6) | | |
| Hf - Te(2)  x2  (b4)$^†$ | 3.062(7) | | |
| Hf - Te(3)  x2  (b5)$^†$ | 2.843(6) | | |

**Table 1.** Crystallographic data for HfTe$_3$. *Isotropic displacement factors were constrained during refinement. $^†$Refer to Fig. 1(c).

**Coexistence of SC and CDW.** Resistivity of non-air-exposed HfTe$_3$ reproducibly exhibited metallic behaviour in the temperature range between 0.3 and 300 K as shown in Fig. 2(a). The residual resistivity ratio (RRR) defined as ρ(275K)/ρ(4K) is ~2.4, which is lower than that of single crystal ZrTe$_3$[11] but is larger than that of polycrystalline-ZrTe$_3$,[24] in which the lower RRR value is thought to arise from strong grain boundary effects. Therefore the influence of grain boundaries is likely to play a role in the reduction of RRR. The inset of Fig. 2(a) shows the temperature derivative of the resistivity dρ/d*T* and reveals a resistivity anomaly at 82 K assumed to be indicative of a CDW formation, where the CDW formation temperature $T_{CDW}$ is defined as the temperature at which dρ/d*T* exhibits a minimum. At $T_{CDW}$ the CDW gap is developed and the resistance anomaly appears owing to a reduction in the density of states at $E_F$ due to the CDW formation. Below 2K, the resistivity showed a sharp drop exhibiting a SC transition at 1.8 K ($T_c^{onset}$) and reached zero ($T_c^{zero}$) at 1.45 K as can be clearly seen in Fig. 2(b). By increasing the applied current, a broadening of the SC transition was observed and it was accompanied by a downward shift in $T_c^{onset}$ and $T_c^{zero}$, whereas the normal state resistivity remains unchanged. The result suggests a weakening in the SC state as well as a decoupling of the Josephson junctions between individual SC grains of the polycrystalline material. *I-V* characteristics measured at $T>T_c$ and $T<T_c$ revealed ohmic and non-ohmic behaviour, respectively. N.B. In the present study, we observe that HfTe$_3$ shows a rapid weakening of its metallic state within minutes of exposure in air (see Supplementary Fig. S1). This is likely the result of an insulating layer (such as tellurium oxides) forming around the individual grains of the polycrystalline material. The results emphasize that if one is to observe the intrinsic properties of HfTe$_3$ any measurements must be conducted in the absence of air.

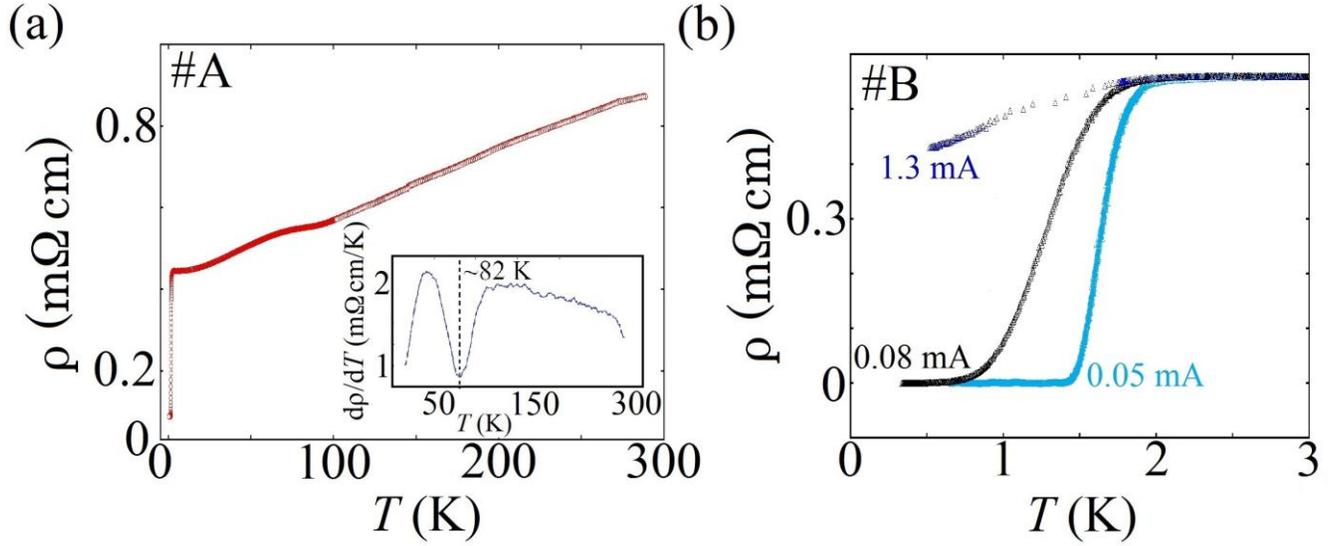

**Figure 2.** (a) Temperature dependence of the resistivity for HfTe$_3$. Data show a hump-like feature at ~80 K together with and SC-like transition at 1.8 K. Inset shows that the resistivity anomaly occurs at approximately 82 K (sample #A). (b) Current dependency of the resistivity of HfTe$_3$ below 3 K. $T_c^{onset}$ is approximately 1.8 K and $T_c^{zero}$ is reached c.a. 1.4 K for a current of 0.05 mA. By an increase in current, both the $T_c^{onset}$ and $T_c^{zero}$ show a shift to lower temperatures (sample #B)

**Behaviour under high-pressure.** By the application of hydrostatic pressure (*P*), the resistivity anomaly gradually shifted to higher temperatures up to ~99 K for *P* approaching 1 GPa as shown in Fig. 3. Similar behaviour has been reported for ZrTe$_3$ where in the case of an application of *P*≤2 GPa the $T_{CDW}$ was increased and the SC suppressed. At *P*≥5 GPa the CDW was fully quenched and gave way to reemergent SC, where $T_c$ increased to ~4.5K when *P*~11 GPa.[15] In addition, in the case of HfTe$_5$, SC appeared by applying *P*~5 GPa and a maximum $T_c$ of 4.8 K was attained by applying at *P*~20 GPa.[25] This suggests the possibility that HfTe$_3$ is likely to follow the same pattern as other members of the group IV-MTe$_x$ alloys. Namely by further application in pressure, it is expected that the $T_{CDW}$ will eventually be suppressed and $T_c$ will be enhanced.

**Electronic structure.** Studies regarding the electronic structure of HfTe$_3$ are limited, but the issue is briefly reported by Felser *et al.* who determined an electronic structure similar to that of ZrTe$_3$, *i.e.* a metallic state resulting from a large contribution of the Te *p*-bands at the Fermi level.[16] These characteristics are supported by later density of states (DOS) calculations.[20, 21] ZrTe$_3$ exhibits a multi-component Fermi surface with contributions from the Te forming quasi 1D electronic sheets at the boundary of the Brillion zone and from the Zr a 3D-hole character sheet around the Γ point. The resultant nesting characteristics at the Fermi surface have been determined to be responsible for the CDW formation in ZrTe$_3$.[16, 26-28] If one considers the iso-structural/electronic relationship between HfTe$_3$ and ZrTe$_3$, it is likely that similar interchain interactions between neighbouring Te(2) and Te(3) atoms (see Fig. 1(c)) play a dominant role in the metallicity of HfTe$_3$[16] which in turn would give rise to the similar Fermi surface with nesting features reported for ZrTe$_3$. However, it cannot be categorically asserted that the observed resistivity anomaly is due to a CDW formation from our results only. As in the case of ZrTe$_3$, it would be necessary to confirm any coincidental low-temperature lattice distortions[29] as well as to observe the features of the Fermi surface around the temperature of the anomaly.[26] However the similarities between HfTe$_3$ and ZrTe$_3$ in the electronic structure as well as the results of the temperature/pressure dependence of ρ are strong indication that the observed resistivity anomaly for HfTe$_3$ is indeed the result of a CDW formation.

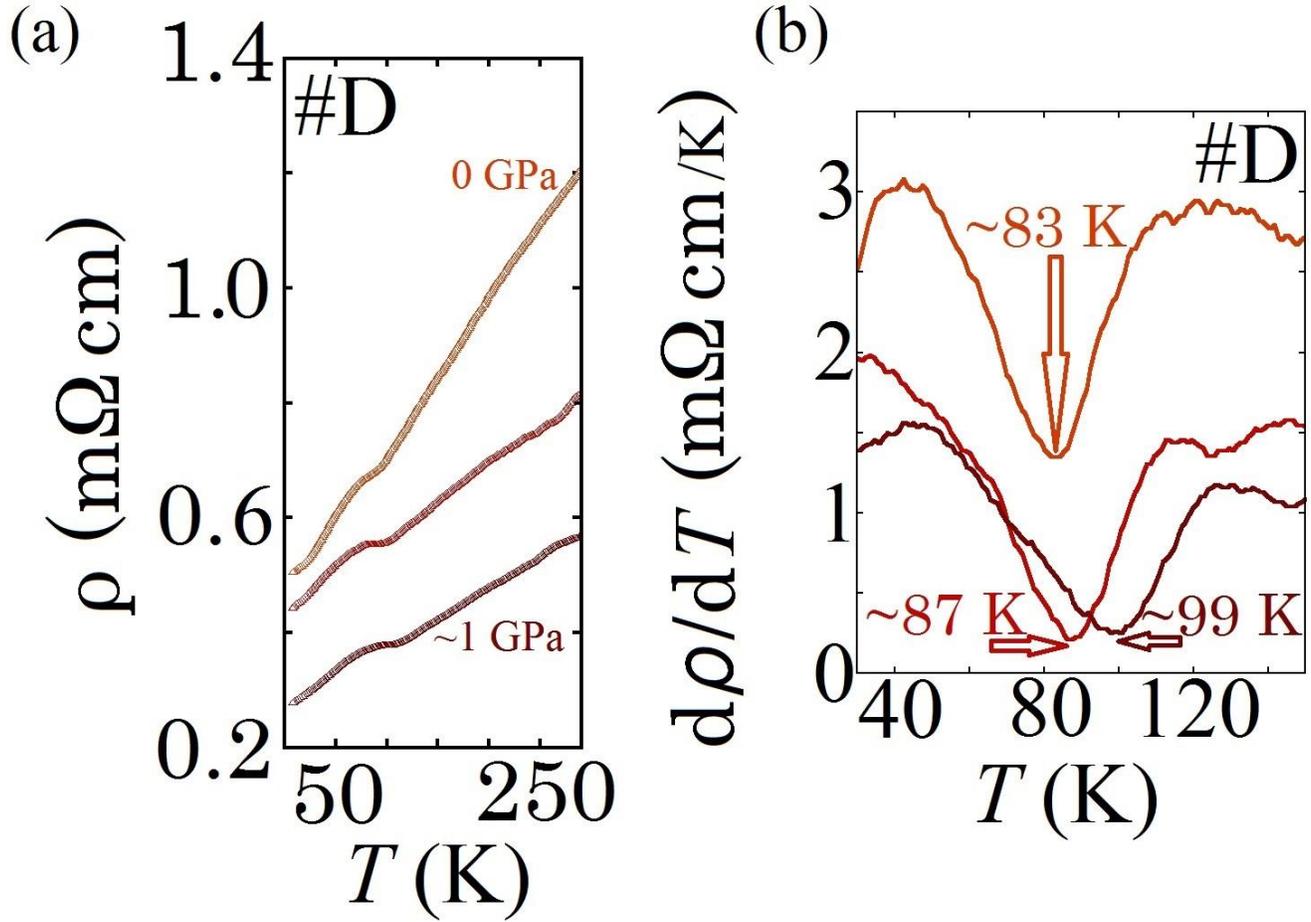

**Figure 3.** (a) Pressure dependence of resistivity at the range of $P$ = 0 - 1 GPa. (b) d$\rho$/d$T$ in the range of 30-150 K as a function of pressure. The minimum of the dip structure shows a shift to higher $T$ with increasing $P$. Color coding between (a) and (b) are matched to indicate the same pressure values (sample #D).

## Conclusion

In summary, we have established a reproducible synthesis method for high-quality polycrystalline HfTe$_3$ and showed that it is an acutely air-sensitive material. By using high-quality HfTe$_3$ we found that the quasi-1D HfTe$_3$ is a novel SC with $T_c$~ 1.4 K, and the SC state coexists with the CDW state which appears at $T_{CDW}$~ 82 K. Furthermore, we provided the first accurate crystallographic data by Rietveld refinement of the PXRD of HfTe$_3$.

## Methods

Single-phase polycrystalline HfTe$_3$ samples have been prepared using standard chemical vapour transport techniques. Ground mixtures of a 1:3 molar ratio of powdered Hf and Te were sealed in silica ampoules under a vacuum of c.a. 3 mTorr using a rotary pump. The ampoules were heated in a box furnace using the reaction procedure described in the results and discussion. To prevent exposure to air, all sample preparation was conducted in an argon filled glovebox.

PXRD was carried out using a Rigaku Smartlab diffractometer in flat plate geometry with a Cu K$_\alpha$ radiation ($\lambda$ = 1.54056 Å). Diffraction data were typically collected for 5° ≤ 2θ ≤ 80° with a 0.01° step size with scan times of 3 hours. Rietveld refinement was performed using the GSAS software package via the EXPGUI interface.[30,31] X-ray fluorescence (XRF) analysis was performed using a JEOL JSX 1000S ElementEye.

Resistivity measurements were performed on cold-pressed pellets using a standard four-terminal setup. Measurements for sample #A were performed between 0.3 and 300 K using an Oxford Instruments 3He cryostat, data were collected by an AC method using a low-noise amplifier and two lock-in amplifiers. Measurements for sample #B were performed by a DC method between 0.3 K and 15 K using a Quantum Design PPMS equipped with an adiabatic demagnetization refrigerator.

The resistivity for Samples #C and #D were also measured by a DC method between 2 and 300 K using a closed cycle helium refrigerator. High-pressure resistivity measurements (up to 1 GPa) were performed using a BeCu/NiCrAl clamped piston-cylinder cell using Daphne 7373 as the fluid pressure transmitting medium with Pb employed as a manometer.

## References

1. Gabovich, A. M., Voitenko, A. I. & Ausloos, M. Charge- and spin-density waves in existing superconductors: competition between cooper pairing and peirels or excitonic instabilities. *Phys. Rep.* **367,** 583-709 (2002).
2. Grüner, G. The dynamics of charge-density waves. *Rev. Mod. Phys.* **60,** 1129-1178 (1988).
3. Wilson, J. A., Di Salvo, F. J. & Mahajan, S. Charge-density waves and superlattices in the metallic layered transition metal dichalcogenides. *Adv. Phys.* **24,** 117-201 (1975).
4. Ido, M., Tsutsumi, K., Sambongi, T. & Mori, N. Pressure dependence of the metal-semiconductor transition in $TaS_3$. *Solid State Comm*. **29,** 399-402 (1979).
5. Wang, Z. *et al*. Charge-density-wave transport above room temperature in a polytype of $NbS_3$. *Phys. Rev. B*. **40,** 11589 (1989).
6. Yamamoto, M. Superconducting properties of $TaSe_3$. *J. Phys. Soc. Jpn.* **45,** 431-438 (1978).
7. Suderow, H., Tissen, V. G., Brison, J.P., Martínez J. C. & Viera S. Pressure induced effects on the fermi surface of superconducting $2H-NbSe_2$. *Phys. Rev. Lett.* **95**, 117006 (2005).
8. Núñez Reguerio, M., Mignot, J. M. & Castello, D. Superconductivity at high pressure in $NbSe_3$. *Europhys. Lett*. **18**, 53-57 (1992).
9. Fang. L. *et al*. Fabrication and superconductivity of $Na_xTaS_2$ crystals. *Phys. Rev. B*. **72,** 014534 (2005).
10. Morosan, E. *et al*. Superconductivity in $Cu_xTiSe_2$. *Nature Phys.* **2**, 544-550 (2006).
11. Takahashi, S., Sambongi, T., Brill, J. W. & Roark, W. Transport and elastic anomalies in $ZrTe_3$. *Solid State Comm*., **49,** 1031-1033 (1984).
12. Zhu, X., Lei, H. & Petrovic, C. Coexistence of bulk superconductivity and charge density wave in $Cu_xZrTe_3$. *Phys. Rev. Lett.* **106,** 246404 (2011).
13. Lei, H., Zhu, X. & Petrovic, C. Raising $T_c$ in charge density wave superconductor $ZrTe_3$ by Ni intercalation. *EPL*. **95,** 17011 (2011).
14. Zhu, X. *et al*. Superconductivity and charge density wave in $ZrTe_{3-x}Se_x$. *Sci. Rep.* **6,** 26974 (2016).
15. Yomo, R., Yamaya, K., Abliz, N., Hedo, M. & Uwatoko, Y. Pressure effect on competition between charge density wave and superconductivity in $ZrTe_3$: appearance of pressure-induced reentrant superconductivity. *Phys. Rev. B*. **71,** 132508 (2005).
16. Felser, C., Finckh. E. W., Kleinke, F., Rocker, F. & Tremel, W. Electronic properties of $ZrTe_3$. *J. Mater. Chem*. **8,** 1787-1798 (1998).
17. Srivastava, S. K. & Avasthi, B. N. Preparation, structure and properties of transition metal trichalcogenides. *J. Mater. Sci.* **27,** 3693-3705 (1992).
18. Brattås, L. & Kjekshus, A. The non-metal rich region of the Hf-Te system. *Acta Chem. Scand.* **25,** 2783-2784 (1971).
19. Brattås, L. & Kjekshus, A. On the properties of compounds with the $ZrTe_3$ type structure. *Acta Chem. Scand.* **26,** 3441-3449 (1972).
20. Abdulsalam, M. & Joubert, D. P. Structural and electronic properties of $MX_3$ (M = Ti, Zr and Hf; = S, Se, Te) from first principles calculations. *Eur. Phys. J. B*. **88,** 177 (2015).
21. Li, M.; Dai, J. & Cheng-Zeng X. Tuning the electronic properties of transition-metal trichalcogenides *via* tensile strain. *Nanoscale*, **7,** 15385-15391 (2015).
22. Wang, Y. *et al*. Spontaneous formation of a superconductor-topological insulator-normal metal layered heterostructure. *Adv. Mater.* **28,** 5013-5017 (2016).
23. Furuseth, S., Brattås, L. & Kjekshus, A. On the crystal structures of $TiS_3$, $ZrS_3$, $ZrSe_3$, $ZrTe_3$, $HfS_3$ and $HfSe_3$. *Acta Chem. Scand.* **29,** 623-631 (1975).
24. Yadav. C.S. & Paulose, P. L. Superconductivity at 5.2 K in $ZrTe_3$ polycrystals and the effect of Cu and Ag intercalation. *J. Phys.: Condens. Matter*. **24,** 235702 (2012).
25. Qi, Y. *et al*. Pressure-driven superconductivity in the transition-metal pentatelluride $HfTe_5$. *Phys. Rev. B*. **94,** 054517 (2016).
26. Kubo, Y. *et al*. Electron momentum density in the low dimensional layered system $ZrTe_3$. *J. Phys. Soc. Jpn.* **76,** 064711 (2007).
27. Stöwe K. & Wagner, F. R. Crystal structure and calculated electronic band structure of $ZrTe_3$. *J. Solid State Chem.* **138,** 160-168 (1998).

28. Hoesch, M. *et al*. Splitting in the fermi surface of ZrTe$_3$: a surface charge density wave system. *Phys. Rev. B.* **80,** 075423 (2009).
29. Eaglesham, D. J., Steeds, J. W. & Wilson, J. A. Electron microscope study of superlattices in ZrTe$_3$. *J. Phys. C: Solid State Phys*. **17,** L697-L698 (1984).
30. Larson, A. C. & von Dreele, R. B. The General Structure Analysis system. Los Alamos National Laboratories, Los Alamos, NM. (2000).
31. Toby, B. H. EXPGUI, a graphical user interface for GSAS. *J. Appl. Crystallogr*. **34,** 210-213 (2001).

## Acknowledgements

The authors would like to thank Akikuni Tomonaga, Keita Miyake and Kensuke Yoshinaga, for their assistance in completing the sample analysis.

## Supplementary information

**Air sensitivity.** We conducted a systematic investigation to analyse the effect air exposure has on the physical properties of HfTe$_3$. HfTe$_3$ shows a rapid weakening of its metallic state within minutes of exposure in air as shown in Fig.S1(a). No change in lattice constants was observed from the PXRD measurement of the 1h air exposed samples as shown in Fig. S1(b). However, PXRD patterns of samples exposed in air for more than one week showed the formation of partially amorphous state as well as degradation in peak intensity. From Fig. S1(c), it is found that the dip in d$\rho$/d$T$ relating to the resistivity anomaly deepened as a function of time exposed in air, but there was no shift in the dip position. On the other hand, HfTe$_3$ stored in an argon atmosphere/vacuum was stable. These results suggest that the insulting behaviour is a non-intrinsic property and a result of an insulating layer forming around the individual grains of the polycrystalline material. It is currently unclear why HfTe$_3$ degrades so easily although the Zr equivalent does not. However it is likely related to the polycrystallinity; almost all ZrTe$_3$ studies have been performed on single-crystalline materials. It had been previously reported that all members of the Zr/Hf-telluride family exhibit oxidative degradation.[1] Single crystal HfTe$_2$[2] as well as powdered ZrTe$_2$[3] are also evidently air-sensitive materials but similar observations have not been reported for ZrTe$_3$. The increased ionicity of Hf in comparison to Zr may also play a role. Further investigation into the thermodynamic stability of the Hf-Te system is required if one is to gain a better understanding of this system.

1. Fjellvåg, H., Furuseth, S., Kjekshus, A. & Rakke, T. Low-temperature oxidative degradation of low-dimensional zirconium and hafnium tellurides. *Solid State Comm*. **63,** 293-297 (1987).

2. Smeggil, J.G. & Bartram, S. The preparation and X ray characterization of HfTe$_{2-x}$, x = 0.061. *J. Solid State Chem.* **5,** 391-394 (1972).

3. Aoki, Y., Sambongi, T., Levy, F. & Berger, H. Thermopower of HfTe$_2$ and ZrTe$_2$. *J. Phys. Soc. Jpn*. **65,** 2590-2593 (1996).

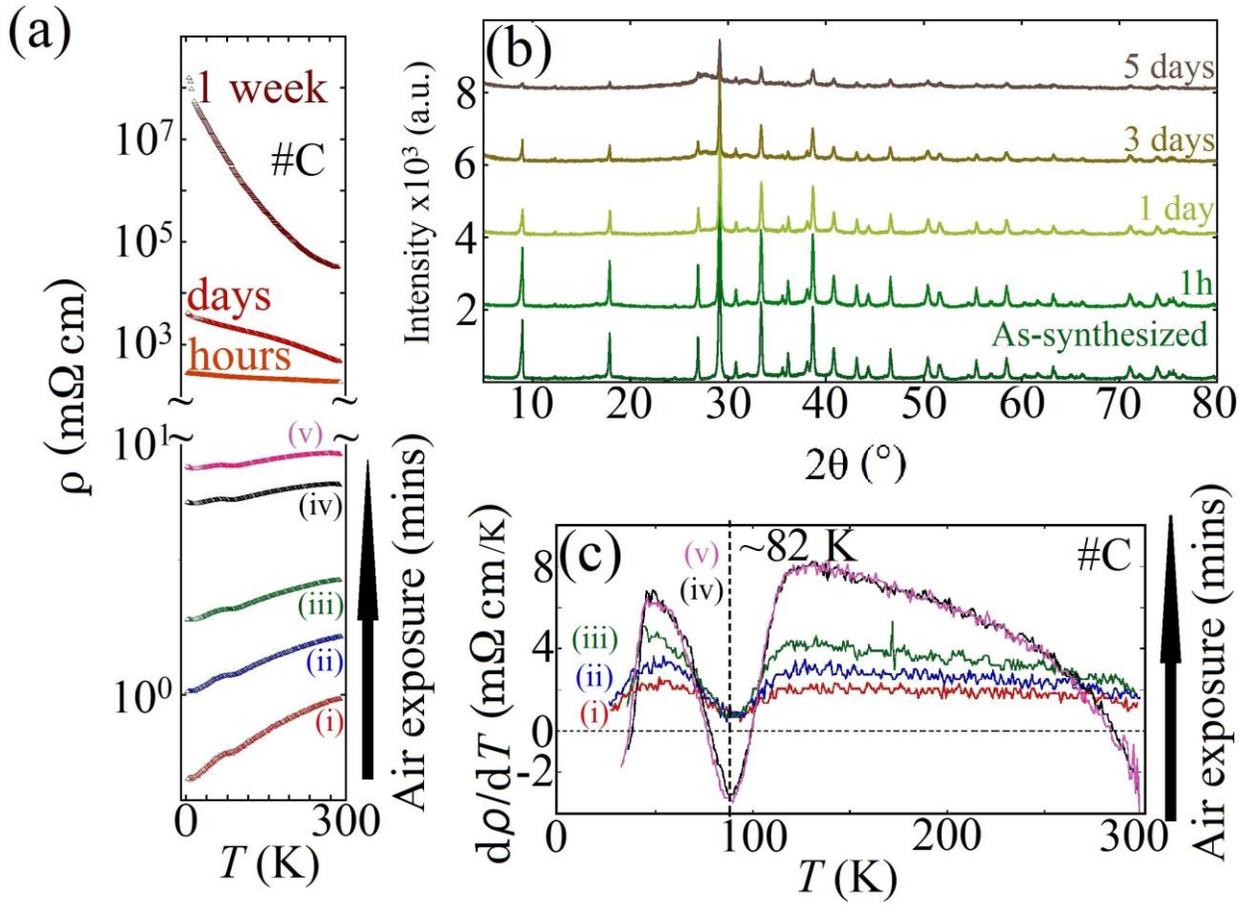

**Figure S1.** (a) Semi-log plot of resistivity of HfTe$_3$ as a function of exposure time in air. (i)- 1min(ii)- 5mins(iii)- 10mins(iv)- 20mins(v)- 40mins(sample #C). (b) PXRD of an air-exposed sample up to five days. Intensity of peaks gradually decreases and an amorphous-like hump feature begins to appear at 20-30° range. (c) Characteristics of d$\rho$/d$T$ for the resistivity data (i) -(v) under air exposure in HfTe$_3$ (sample #C).